# Formation of Charon's Red Poles From Seasonally Cold-Trapped Volatiles

Grundy, W.M.[1*], Cruikshank, D.P.[2], Gladstone, G.R.[3], Howett, C.J.A.[4], Lauer, T.R.[5], Spencer, J.R.[4], Summers, M.E.[6], Buie, M.W.[4], Earle, A.M.[7], Ennico, K.[2], Parker, J.Wm.[4], Porter, S.B.[4], Singer, K.N.[4], Stern, S.A.[4], Verbiscer, A.J.[8], Beyer, R.A.[10/2], Binzel, R.P.[7], Buratti, B.J.[9], Cook, J.C.[4], Dalle Ore, C.M.[10/2], Olkin, C.B.[4], Parker, A.H.[4], Protopapa, S.[11], Quirico, E.[12], Retherford, K.D.[3], Robbins, S.J.[4], Schmitt, B.[12], Stansberry, J.A.[13], Umurhan, O.M.[2], Weaver, H.A.[14], Young, L.A.[4], Zangari, A.M.[4], Bray, V.J.[15], Cheng, A.F.[14], McKinnon, W.B.[16], McNutt Jr., R.L.[14], Moore, J.M.[2], Nimmo, F.[17], Reuter, D.C.[18], Schenk, P.M.[19], and the New Horizons Science Team.

[1]Lowell Observatory, Flagstaff AZ (w.grundy@lowell.edu), [2]NASA Ames Research Center, Moffett Field CA, [3]Southwest Research Institute, San Antonio TX, [4]Southwest Research Institute, Boulder CO, [5]National Optical Astronomy Observatory, Tucson AZ, [6]George Mason University, Fairfax VA, [7]Massachussetts Institute of Technology, Cambridge MA, [8]University of Virginia, Charlotteville VA, [9]NASA Jet Propulsion Laboratory, La Cañada Flintridge CA, [10]Carl Sagan Center at the SETI Institute, Mountain View CA, [11]University of Maryland, College Park MD, [12]Université Grenoble Alpes, CNRS, IPAG, Grenoble France, [13]Space Telescope Science Institute, Baltimore MD, [14]Johns Hopkins University Applied Physics Laboratory, Columbia MD, [15]University of Arizona Lunar and Planetary Laboratory, Tucson AZ, [16]Washington University in St. Louis, St. Louis MO, [17]University of California, Santa Cruz CA, [18]NASA Goddard Space Flight Center, Greenbelt MD, [19]Lunar and Planetary Institute, Houston TX.
[*]Corresponding author.

**A unique feature of Pluto's large satellite Charon is its dark red northern polar cap[1]. Similar colours on Pluto's surface have been attributed[2] to organic macromolecules produced by energetic radiation processing of hydrocarbons. The polar location of this material on Charon implicates the temperature extremes that result from Charon's high obliquity and long seasons. The escape of Pluto's atmosphere provides a potential feed stock for production of complex chemistry[3,4]. Gas from Pluto that is transiently cold-trapped and processed at Charon's winter pole was proposed[1,2] as an explanation on the basis of an image of Charon's northern hemisphere, but not modelled quantitatively. Here we report images of the southern hemisphere illuminated by Pluto-shine and also images taken during the approach phase showing the northern polar cap over a range of longitudes. We model the surface thermal environment on Charon, the supply and temporary cold-trapping of material escaping from Pluto, and, while cold-trapped, its photolytic processing into more complex and less volatile molecules. The model results are consistent with the proposed mechanism producing the observed colour pattern on Charon.**

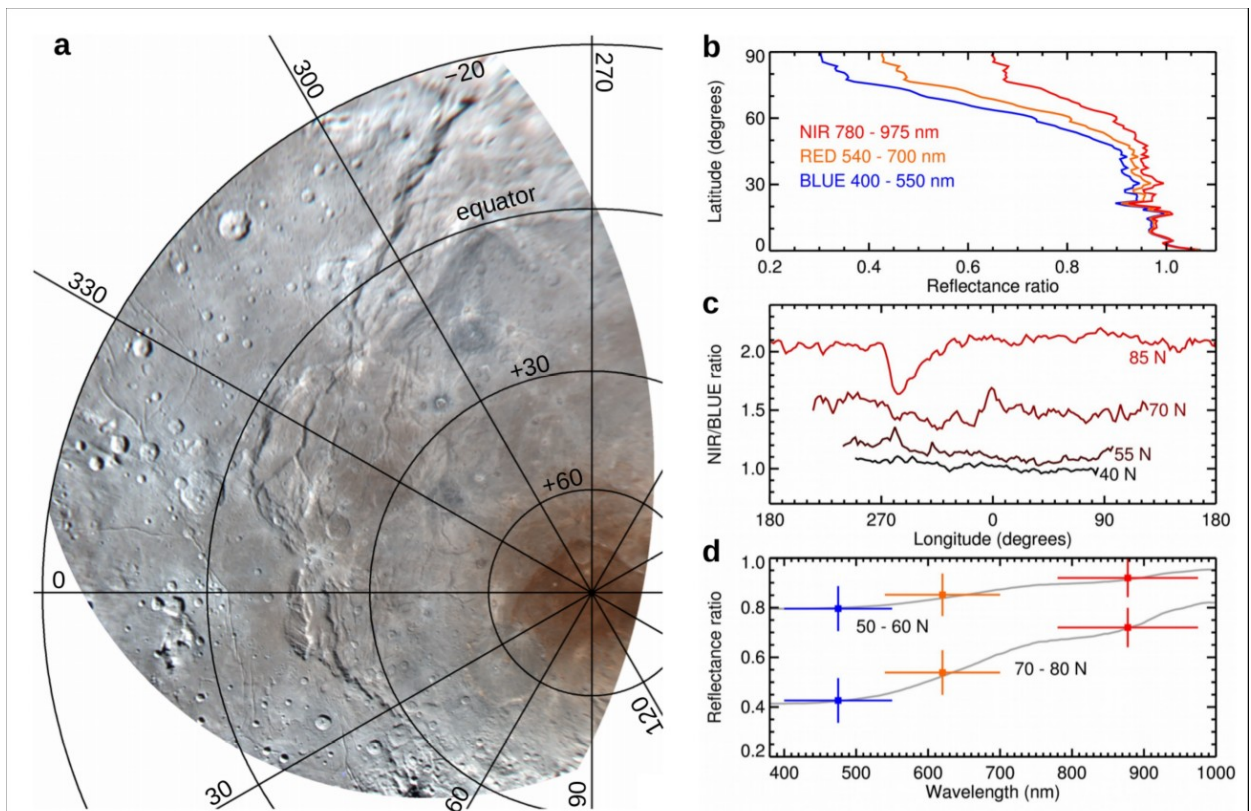

**Figure 1. Charon's red northern pole. a.** Polar stereographic projection with Ralph's BLUE, RED, and NIR filter images displayed in blue, green, and red colour channels, respectively, relative to a Hapke photometric model (see Methods). **b.** Latitude dependence of reflectance relative to the photometric model. **c.** Longitudinal dependence of the NIR/BLUE colour ratio. **d.** Wavelength dependence at two latitudes (coloured points) compared with spectral models of a laboratory tholin plus a neutral material (grey curves). Vertical bars indicate standard deviation within each latitude bin. Horizontal bars indicate filter widths.

Fig. 1a shows Charon's northern polar cap. The image combines the BLUE (400-550 nm), RED (540-700 nm), and NIR (780-975 nm) channels from the Multispectral Visible Imaging Camera (MVIC), part of New Horizons' Ralph remote sensing package[5,6]. Fig. 1b shows how reflectance in the three filters varies with latitude, averaged over longitudes shown in Fig. 1a. Charon's surface gets darker and more red toward higher latitudes. Ratios of NIR/RED and RED/BLUE show similar latitude dependence to one another[2], suggesting a single pigment material with increasing abundance toward the pole. Additional approach images are shown in Extended Data Fig. 1.

Longitudinal variability in the NIR/BLUE colour ratio is shown in Fig. 1c. At higher latitudes, colours are redder across all longitudes observed, although the trend is not perfectly uniform. Deviations may be related to local variations in topography or other parameters. The red coloration is interrupted by a few impact craters in the several km size range. Impacts that size occur rarely, likely much less frequently than once per million years[7], so their existence implies the red material must accumulate slowly (see Methods).

Infrared spectroscopy also favours slow accumulation. Spectra of Charon's pole are dominated by $H_2O$ ice absorptions similar to spectra of lower latitude regions[2]. The data are consistent with up to ~10% tholin mixed with $H_2O$ ice at the ~mm depths probed by the infrared observations, so tholin deposition must occur slowly enough that $H_2O$ resupply or upward mixing of $H_2O$ by impact gardening can compete.

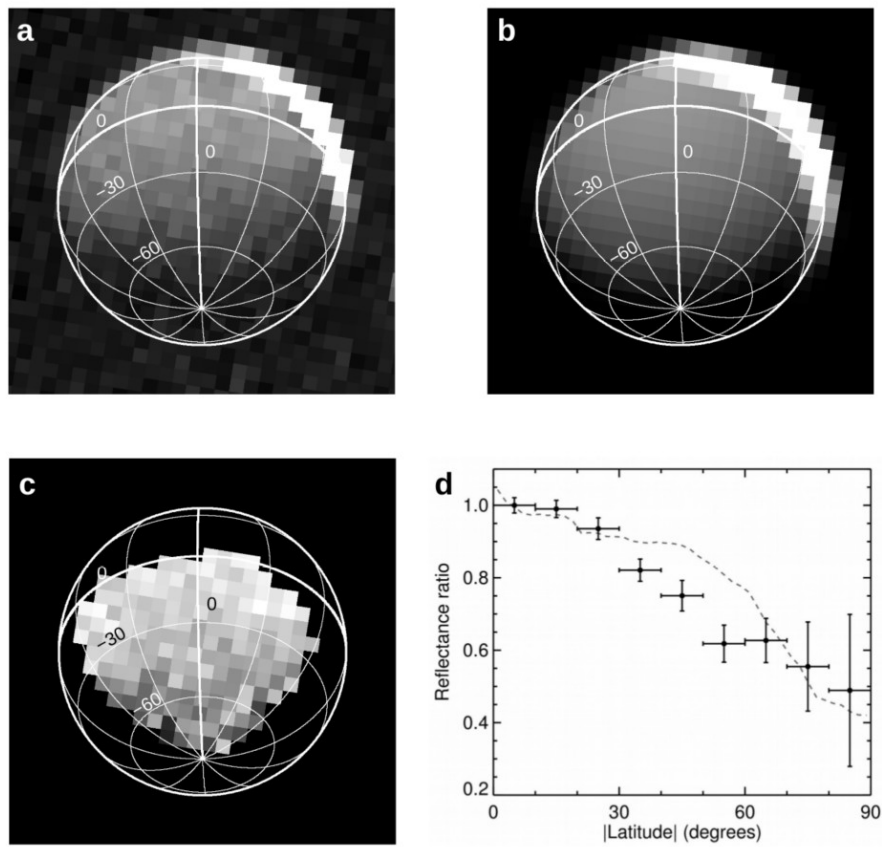

**Figure 2. Winter pole in Pluto-shine. a.** Stack of 99 images showing a bright, sunlit crescent and fainter reflected light from Pluto. North, defined by the angular momentum vector, is up. **b.** Photometric model assuming a uniform albedo (see Methods). **c.** Observation/model ratio, showing southern high latitudes are dark relative to equatorial latitudes. **d.** Latitude dependence of the ratio. Points are for the Pluto-shine illuminated southern hemisphere. Horizontal bars indicate latitude bin width. Vertical bars show the standard deviation of the mean within each latitude bin. The dashed curve is for the sunlit northern hemisphere (see Methods).

Charon's south pole is currently in winter night. New Horizons observed the night-side with the LOng Range Reconnaissance Imager (LORRI[8]) ~2.6 days after closest approach, illuminated by "Pluto shine." The images reveal a decreasing brightness toward the pole (Fig. 2 and Extended Data Fig. 2) that cannot be attributed to declining illumination alone, but requires decreased albedo of the pole relative to equatorial latitudes (see Methods). The sunlit northern hemisphere in LORRI approach images shows a comparable albedo decline toward that pole (Fig. 2d).

Charon's surface temperature responds to solar forcing on diurnal (6.39 Earth days) and annual (248 Earth years) time scales. The high obliquity (currently 119°) causes polar latitudes to experience long periods of continuous darkness, during which they become extremely cold. Complicating the situation, the eccentricity of Pluto's heliocentric orbit, currently 0.253, results in a factor of 2.8 difference in intensity of sunlight between perihelion and aphelion. To assess the thermal history of Charon's surface, we ran thermophysical models[9] tracking diurnal and annual vertical heat flow into and out of Charon's surface at different latitudes, accounting for the present day orbital parameters[10] (see Methods). Fig. 3a shows model thermal histories for four different northern hemisphere latitudes over the past few centuries for a nominal thermal inertia of 10 J m$^{-2}$ K$^{-1}$ s$^{-½}$ consistent with estimates of Charon's diurnal thermal inertia from Herschel Space Telescope observations[11]. Models for bracketing thermal inertias 2.5 and 40 J m$^{-2}$ K$^{-1}$ s$^{-½}$ are shown in Extended Data Fig. 3, spanning the range reported from icy satellites[12] to Kuiper belt objects[13]. Charon's north pole experienced more than a century of continuous, extremely low temperatures from the late 1800s through the spring equinox in 1989. Lower latitudes experienced briefer periods of continuous extreme cold, and also reached less extreme minimum temperatures.

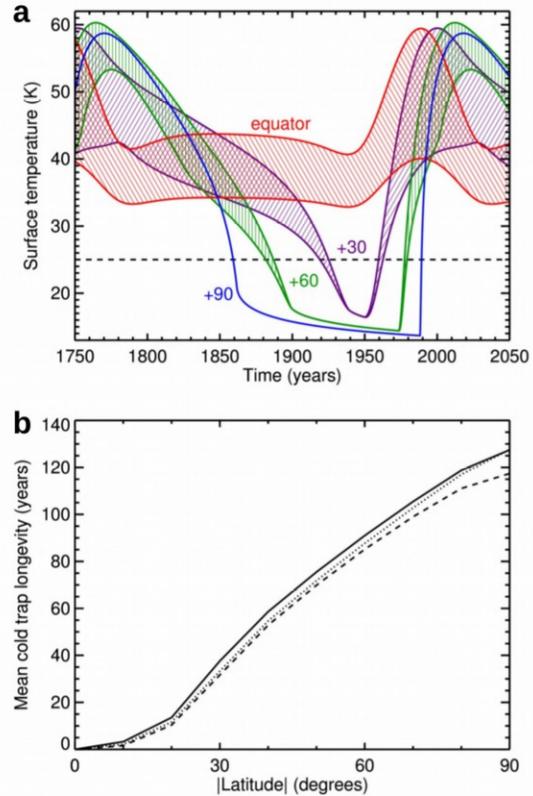

**Figure 3. Thermal environment. a.** Model surface temperature history for equatorial and three northern latitudes on Charon from 1750 through 2050, for thermal inertia 10 J m$^{-2}$ K$^{-1}$ s$^{-½}$. Envelopes for each latitude indicate diurnal minimum and maximum temperatures. The dashed line is the 25 K threshold below which CH$_4$ is cold-trapped. **b.** Longest continuous duration below that temperature each Charon year, averaged over thermal models spanning the past 3 Myr. Dotted, solid, and dashed curves are for thermal inertias 2.5, 10, and 40 J m$^{-2}$ K$^{-1}$ s$^{-½}$, respectively.

The latest Charon year shown in Fig. 3a provides an incomplete picture of Charon's long term thermal history, because the pole precesses and the longitude of perihelion regresses[10,14], both on 3 × 10$^6$ year time scales. For an idea of Charon's longer term thermal history, we ran thermal models for prior Charon years, selecting one every 4 × 10$^5$ years over the past 3 × 10$^6$ years. Durations of continuous periods colder than a 25 K threshold temperature for cold trapping (derived below) were averaged over the modelled epochs and also over the northern and southern hemispheres. These are shown as a function of latitude in Fig. 3b. Higher latitudes experience longer periods below the threshold temperature.

CH$_4$ and N$_2$ currently escape from Pluto at rates of 5 × 10$^{25}$ and 1 × 10$^{23}$ molecules s$^{-1}$, respectively, as estimated from New Horizons data[4], with the Pluto exobase located at ~2.5 Pluto radii (i.e., at an altitude of ~1780 km)[15]. These are different from conditions assumed in a pre-encounter study by Tucker et al.[3], who investigated the transfer of Pluto's escaping atmosphere to Charon including condensation on the winter pole. Their model was based on an N$_2$ escape rate from Pluto of 2.3 × 10$^{27}$ molecules s$^{-1}$, of which 5.7 × 10$^{25}$ molecules s$^{-1}$ (i.e., ~2.5%) encountered Charon, leading to 0.2 μm decade$^{-1}$ N$_2$-ice deposition on the winter pole. More recent simulations[16] report comparable arrival fractions. The long term temporal variability of the escape rate is not known.

Assuming Charon intercepts 2.5% of the New Horizons determined $CH_4$ escape flow, that would correspond to a globally averaged arrival rate of $2.7 \times 10^{11}$ molecules $m^{-2}$ $s^{-1}$ at Charon. We estimate that most of the $CH_4$ remains at Charon long enough to find its way to the winter pole (see Methods). If it accumulates within 45º of the winter pole (a typical polar size, see Fig. 3b), it would produce a ~0.3 µm layer of $CH_4$ ice at each pole of Charon during the winter portion of each Pluto year. The accumulation would presumably vary with latitude, being thicker toward the pole.

Condensation on a surface depends on the temperature and vapour pressure[17]. The pressure at Charon's surface depends on the area of the polar cold trap. For a 45º radius pole, it can be estimated as $1 \times 10^{-11}$ Pa. This pressure corresponds to an equilibrium vapour pressure of $CH_4$ ice at a temperature of 25 K[18]. Where the surface temperature is colder, $CH_4$ will tend to freeze out as ice.

Methane escaping from Pluto's atmosphere is accompanied by other minor species. These include about 0.2% $N_2$, various $C_2$ hydrocarbons (at a few 10s of ppm), and radicals such as $CH_3$ (at ~100 ppm). $N_2$ is more volatile and thus requires lower temperatures to be cold trapped, so it should freeze onto Charon's surface over a smaller range of latitudes and times of year than $CH_4$ does. Heavier hydrocarbons can condense anywhere on Charon, but would not produce enough of a deposit to be visible.

Our hypothesis requires energetic radiation to process the seasonally cold-trapped $CH_4$. It is only frozen on Charon's surface during the polar winter night, so it must be processed rapidly, on the timescale of a century, and only by radiation impinging on the night side. It need not be fully converted into macromolecular solids such as tholins on such a short timescale, only into molecules that are sufficiently non-volatile to remain on the surface after the pole re-emerges into sunlight and warms back up. Charon's surface is subject to a variety of energetic radiation sources, including UV photons, solar wind charged particles, interstellar pickup ions, and galactic cosmic rays[19,20]. The most important night side source of energetic radiation appears to be solar ultraviolet Lyman alpha photons (Lyα, 10.2 eV) that have been scattered by the interplanetary medium, with a flux of $3.5 \times 10^{11}$ Lyα photons $m^{-2}$ $s^{-1}$ on the night side. For $2.7 \times 10^{11}$ molecules $m^{-2}$ $s^{-1}$ arriving at Charon, concentration by cold trapping in a 45° cold pole (~1/7 of Charon's surface area) results in accumulation of ~3 nm of ice per Earth year, of which we estimate 21% is photolyzed (see Methods). There would likely also be some loss to sputtering[21]. $N_2$ ice is unaffected by Lyα[22] but some could be processed by cosmic rays.

When the winter pole re-emerges into sunlight, $CH_4$ and $N_2$ sublimate away rapidly, but heavier, less volatile products remain behind. Assuming the mass density of these photolytic products is double that of $CH_4$ ice, around ~40 nm would accumulate per Pluto winter, or 0.16 mm per million Earth years. They will be exposed to additional sources of energetic radiation including UV and EUV photon radiation directly from the Sun, driving further photolytic chemistry[23] as well as sputtering erosion. At Pluto and Charon's mean heliocentric distance of 39 AU, the solar Lyα energy flux is $1.9 \times 10^{13}$ eV m$^{-2}$ s$^{-1}$, and that of the EUV (>12.4 eV) is $8.7 \times 10^{11}$ eV m$^{-2}$ s$^{-1}$. The charged particle flux from the ambient solar wind, coronal mass ejections, and interstellar pickup ions is highly variable, and has energies ranging from a few eV to MeV. The production efficiency of tholin is not known for Charon's circumstances, but if it were 50%, that would translate to ~30 cm of tholin produced at Charon's poles over 4 billion years.

Strongly coloured tholins have been made in the laboratory from a variety of $CH_4$ and $N_2$ ice mixtures using diverse radiation sources from charged particles through UV photons[24,25]. Upon ~10× further irradiation[20,25], experiments that initially produced a red-orange tholin generally go on to produce colourless and much darker material, indicating carbonization (graphitization). The fact that Charon's pole is not completely blackened requires a balance between production and further processing of tholins and processes such as in-falling $H_2O$ ice dust or ejecta or micrometeorite impact gardening that would mix them into the uppermost mm to few meters of Charon's $H_2O$ regolith. Two distinct sources could contribute. Low-velocity dust from Pluto's small satellites is estimated to produce a 3 cm coating over the course of 4 Gyr[26] with a composition likely dominated by $H_2O$ ice. This is an order of magnitude slower than our estimated tholin production, although it could be augmented by ejecta from large impacts elsewhere on Charon. Higher velocity Kuiper belt debris impact rates are poorly constrained, but models based on lunar rates[27,28] imply gardening to cm depth on a timescale of $10^7$ years (see Methods), during which time we estimate about a mm of tholin accumulates. By diluting the accumulating tholin in local substrate material, this shallow gardening may explain the enduring brightness of a few relatively recent craters where accumulating tholin is mixed into more neutral coloured, $H_2O$-rich ejecta, rather than being mixed into the already-darkened substrate, as elsewhere.

The distribution of dark, reddish material around Charon's northern pole is notable for its generally symmetric distribution in longitude and its gradual increase with latitude, although there are local irregularities associated with craters, topographic features, and perhaps sub-surface variations in thermal properties. These characteristics and the existence of an albedo feature around the southern pole with a similar latitude dependence are consistent with our hypothesis that the combination of Pluto's escaping atmosphere and Charon's long, cold winters enables $CH_4$ to be seasonally cold trapped at high latitudes, where some is photolytically processed into heavier molecules that are subsequently converted to reddish tholin-like materials. The symmetry argues against recent polar wander on Charon. Could the process occur elsewhere? Nix has a reddish spot[29], but it and the other small satellites orbit farther from Pluto and have much lower masses, making the process less efficient for them (see Methods).

## Acknowledgements

This work was supported by NASA's New Horizons Project.  E.Q., B.S., and S.Philippe. acknowledge the Centre National d'Etudes Spatiales (CNES) for its financial support through its "Système Solaire" program.

## Methods

### Colour images

The Charon colour image shown in Fig. 1a was obtained on 2015 July 14 at 10:42 UT. The spacecraft was 73,000 km from Charon, resulting in an MVIC image scale of 1.4 km/pixel. The Mission Elapsed Time (MET) unique label of this observation was 0299176432. Howett et al. (2016) describe calibration details[6]. The original image is shown in Extended data Fig. 3c. Fig. 1a shows it reprojected to a polar stereographic projection and photometrically corrected by dividing each pixel by a Hapke[30] model computed for the same illumination and viewing geometry. That model had uniform parameters across the scene: single scattering albedo $w = 0.9$, single scattering phase function $P(g) = 0.8$, backscattering amplitude $B_0 = 0.6$, porosity parameter $h = 0.0044$, and macroscopic roughness $\bar{\theta} = 20°$ (parameters from Buie et al. 2010[31]). Fig. 1b shows the latitude dependence of the image divided by the Hapke model, averaged over longitudes shown in Fig. 1a and normalized to unity at the equator. Similar latitude dependence of Charon's normal albedo has been reported from LORRI observations[32]. Longitude and wavelength dependence are shown in Figs. 1c and 1d. Additional approach images are shown in Extended Data, without reprojection or photometric correction.

### Pluto-shine observations

New Horizons obtained 219 LORRI images of Charon's night side early on 2015 July 17 UT, from a range of approximately 3 million km. The spacecraft's orientation is controlled by hydrazine thrusters, so full resolution LORRI exposures must be short to minimize smear as pointing bounces around within a deadband. Longer exposures of 0.2 seconds were enabled by 4×4 pixel on-chip binning, resulting in a total integration of 44 s. The images were grouped into two sets, the first comprising 99 images with MET labels 0299398349 through 0299398716 acquired at a mean observation time of 00:26 UT. The second set had 120 images with MET labels 0299405549 through 0299405916 and a mean time of 02:26 UT. The two sets had image scales of 60 and 62 km/pixel, respectively. As LORRI was pointed close to the sun to observe the night side of Charon, the images were affected by scattered light patterns modified by small pointing variations among the images. The scattered light variability in each image set was modelled using Principal Component Analysis on the complementary set, after masking Charon and bright stars from the images. The eigenimages in each set were used to subtract off the scattered light contribution from the other set. The images were then co-registered at the sub-pixel level and combined into two stacks, the first of which is shown in Fig. 2a, the second in Extended Data Fig. 2a.

We modelled these observations by simulating Charon as a sphere approximated by 20480 triangular facets. The bidirectional reflectance behaviour of each facet was represented with a Hapke model, using the same parameters as above. The Sun was treated as a point source and the distribution of light from Pluto was obtained by assuming it to be a Lambertian sphere with the appropriate size, location, and illumination geometry. Being larger than Charon, Pluto casts some light onto Charon's pole, albeit obliquely.  Comparison between data and model are shown for the first of the two image stacks in Fig. 2 and for the other stack in Extended Data Fig. 2.

To compare Charon's southern pole with the northern one as seen with the same instrument, we also ran a Hapke model with the same parameters for a LORRI full resolution sunlit approach image obtained 2015 July 14 at 02:44 UT, from a range of 470,000 km, with image scale 2.3 km/pixel.  The MET label for this image was 299147776.  It and the corresponding model are shown in Extended Data Figs. 2d and 2e, and the latitude-dependent ratio is included as a dashed curve in Fig. 2d and Extended Data Fig. 2f, normalized to unity over the mean from 0° to 10° N latitude.  MVIC colour observations were not sufficiently sensitive to detect Pluto-shine on Charon's night side.

## Thermal model

A standard one dimensional finite element model[9] was used to account for vertical heat flow within each element of Charon's surface in response to diurnally and seasonally varying insolation[10].  Diurnal and annual time scales differ greatly, so we represented Charon's surface with a large number of layers (400) and broke its year into a large number of time steps ($2 \times 10^6$, about an hour per time step).  We assumed a uniform bolometric bond albedo $A_B = 0.3$ and emissivity $\varepsilon = 0.9$, ignoring the lower albedos of Charon's poles.  Diurnal thermal inertia $\Gamma$ values were varied from 2.5 to 40 J m$^{-2}$ K$^{-1}$ s$^{-\frac{1}{2}}$, spanning the range reported for Charon and Kuiper belt objects from Spitzer and Herschel observations[11,13] to icy saturnian satellites observed by Cassini[12].  The density and heat capacity of $H_2O$ ice are fixed, so low thermal inertias require inefficient conduction between ice grains.  Conduction can rise with compaction at depth, leading to higher seasonal thermal inertias.  Simulations with higher conductivities up to that of monolithic $H_2O$ ice[33] at depths as shallow as few 10s of cm below the surface can raise polar winter temperatures by a few K without much altering diurnal temperature variations.  The only energy source was assumed to be sunlight, 870 mW m$^{-2}$ at Pluto's mean heliocentric distance.  We did not include radiogenic heating.  If Charon's ~60% rock fraction[34] has chondritic radionuclide abundances, heat from their decay would contribute an additional 1.5 mW m$^{-2}$, preventing the temperature from falling below about 13 K at present, and higher in the distant past.  Fig. 3a shows model surface temperature history over the course of three centuries for Charon's northern hemisphere assuming an intermediate thermal inertia $\Gamma = 10$ J m$^{-2}$ K$^{-1}$ s$^{-\frac{1}{2}}$, no radiogenic heat, and no high-conductivity subsurface layer.  High and low $\Gamma$ bounding cases are shown in Extended Data Fig. 3.  Diurnal variations are much greater for small $\Gamma$, as is the range of seasonal surface temperature extremes.  For all of these scenarios, Charon's high latitudes experience multi-decade episodes sufficiently cold to cold-trap $CH_4$ as ice.

As the spin pole of the system precesses and the longitude of perihelion of its heliocentric orbit regresses[10,14], the insolation patterns evolve, modifying the surface temperature history over the course of a Pluto year.  This variability is accommodated in Fig. 3b by averaging the longest continuous period below 25 K during a Pluto year over both hemispheres and over 3 million years of orbital history.

## Loss mechanisms and cold trapping

$CH_4$ molecules can be lost from Charon's surface environment via several mechanisms. First we consider thermal escape to space. For a Maxwell-Boltzmann distribution at 60 K, corresponding to the highest temperatures reached on Charon's summer hemisphere (see Fig. 3a), about 1% of $CH_4$ molecules exceed Charon's 590 m s$^{-1}$ escape velocity. The majority that do not escape hop on ballistic trajectories until they achieve escape velocity or encounter the cold pole and stick. A typical latitude boundary of the cold pole from Fig. 3b is 45°. A cold pole above that latitude occupies about 1/7 of Charon's surface. If each ballistic hop is assumed to arrive in a completely new random location, it would take an average of 7 random hops to encounter the cold pole at a time when it that size. With 1% lost to space per hop, 93% would make it to the pole, leading to an accumulation rate there of $2 \times 10^{12}$ molecules m$^{-2}$ s$^{-1}$. For $CH_4$ ice density[35] of 516 kg m$^{-3}$, the deposition rate is $9 \times 10^{-17}$ m s$^{-1}$ which adds up to about 0.3 µm at the pole over one Charon winter. Since the accumulation time is strongly latitude-dependent, the resulting ice distribution would be, too.

Another potential loss mechanism is ionization. If a molecule is ionized, it becomes coupled to the magnetic field in the solar wind and is thus swept from the system. The cross section of a $CH_4$ gas molecule to Lyα radiation is estimated[22] as $1.8 \times 10^{-21}$ m$^2$. The solar Lyα flux at Charon's mean heliocentric distance of 39 AU is $1.9 \times 10^{12}$ cm$^{-2}$ s$^{-1}$, so the probability of photoionization is about 0.1 per Earth year. At typical thermal speeds of 200 to 300 m s$^{-1}$, $CH_4$ molecules can traverse Charon's diameter in less than 2 hours, so the probability of photoionization during 7 ballistic hops is low. Likewise, little $CH_4$ would be photoionized between its escape from Pluto and arrival at Charon, considering that Pluto's exobase temperature is ~70 K, so molecules escaping at the tail of the Maxwellian velocity distribution would reach Charon in just a few tens of hours.

## Surface pressure

An estimate of the surface pressure at Charon can be obtained by assuming steady state, with arriving $CH_4$ molecules undergoing an average of 7 ballistic hops before being lost to space or cold-trapped on a 45° radius cold pole, so each element of Charon's surface experiences 7× the globally averaged $CH_4$ arrival flux of $2.7 \times 10^7$ molecules cm$^{-2}$ s$^{-1}$ at speeds consistent with the Boltzmann velocity distribution, resulting in a momentum flux or pressure of roughly $1 \times 10^{-11}$ Pa, three orders of magnitude smaller than the 3σ upper limit from New Horizons[36]. When the cold pole is smaller, the pressure increases because more hops occur, and the reverse is true when the cold pole is larger, so this pressure is very approximate. However, it enables us to estimate the threshold temperature of 25 K for cold trapping $CH_4$, via its vapor pressure[18]. Assuming the radius of a $CH_4$ molecule is $2 \times 10^{-10}$ m, this pressure corresponds to a mean free path >100 Charon radii, so collisions between $CH_4$ gas molecules can be ignored.

## Photolysis of $CH_4$ ice

The interplanetary medium scatters Lyman alpha (Lyα) photons from the Sun. The flux from this source observed during the New Horizons flyby[4] was about 50% larger than predicted[37], or about 140 rayleigh averaged over the anti-Sun hemisphere that illuminates the winter pole, equating to a flux of $3.5 \times 10^{11}$ Lyα photons m$^{-2}$ s$^{-1}$. This appears to be the largest source of night-side UV illumination at wavelengths between 10 and 133 nm that efficiently break bonds in the $CH_4$ molecule[22,38]. Integrating over these wavelengths for the 1000 stars with largest apparent UV fluxes using IUE spectra and Kurucz[39] models and dividing by two to account for the visible 2π steradians of sky contributes an additional $1.1 \times 10^{10}$ photons m$^{-2}$ s$^{-1}$. The Lambert absorption coefficient of $CH_4$ ice at Lyα wavelength (1216 Å)[38] is 19 µm$^{-1}$ so a 53 nm thick $CH_4$ ice coating has optical depth unity. The Lyα cross section of individual $CH_4$ ice molecules has been reported[22] as $1.4 \times 10^{-21}$ m$^2$, enabling estimation of the probability of photolysis of an unshielded $CH_4$ ice molecule on Charon's winter pole as 1.5% per Earth year. For $2.7 \times 10^{11}$ molecules m$^{-2}$ s$^{-1}$ arriving at Charon, concentration by cold trapping in ~1/7 of Charon's surface area results in accumulation of ~3 nm of $CH_4$ ice per Earth year, or a layer that is optically thick to Lyα in 18 years. Only the skin reachable by UV photons can be photolyzed. For steady state at this deposition rate, around 26% of the $CH_4$ would be processed before becoming buried beneath enough $CH_4$ ice to shield it from further UV photolysis. Over a century-long accumulation cycle at this rate, about 21% is processed.

## Impacts and gardening

Only one of the bright-ray craters in Charon's polar region appears to be resolved. Its darker, inner region could mark the rim, or it could be ejecta since many of Charon's other craters feature dark inner ejecta. If the dark/bright boundary is the rim, the diameter of the crater itself is ~5 to 6 km; otherwise, scaling from fully resolved craters, the diameter could be as small as ~1.5 km. In addition to uncertainty about crater size, the impactor flux in the outer solar system for objects below ~100 km is not well known. Using the Greenstreet et al. "knee" model[40,41] (consistent with the size distribution of larger craters on Pluto and Charon[7]), one ~5 to 6 km or larger crater should typically form somewhere on Charon every ~1 million years. The probability of occurrence in dark polar material is lower, in proportion to its smaller surface area. If the crater is smaller than 5 to 6 km, the occurrence frequency would be higher, but the Greenstreet et al. model predicts many more craters smaller than 10 km than are observed on Pluto or Charon[7], suggesting the timescale for formation of the observed craters could be considerably longer than a million years.

To date, no detailed modelling has been conducted of impact gardening of outer solar system objects. Two studies of lunar impact gardening found similar results for $10^7$ year timescales[27,42]. Gault et al.[27] estimated a 50% probability that material has been excavated down to a depth of 1 cm over that interval, while Arnold[42] found all study points in a grid were excavated at least once down to a depth of ~1 cm. The two studies diverged for longer timescales, with Arnold deriving a disturbed depth of 10 cm in $10^8$ years, and the Gault model taking $10^9$ years to achieve the same disturbed depth.

The rate of gardening in the lunar example is calculated for impactors of all sizes, from the largest objects to hit the Moon, down to micrometeorites and dust-sized particles. Extrapolation to the outer solar system requires knowledge of the impactor flux, which, as described above, is poorly constrained. The lunar flux used by Gault is larger than the best estimate from Greenstreet et al. for Charon for impactors larger than ~0.1 mm by up to several orders of magnitude, but smaller than the extrapolated Greenstreet et al. model for smaller impactors by similar amounts. The dust flux (particles below 1 cm in diameter) estimated from the New Horizons Student Dust Counter[43] is 4–6 orders of magnitude below the Greenstreet et al. model. Extrapolating from that model to much smaller sizes or from the New Horizons dust measurement to much larger sizes is highly uncertain, illustrating the broad range of potential impactor flux.

## Other instances

A natural question to ask is whether our hypothesized mechanism for production of Charon's dark, red poles should produce similar deposits elsewhere. Unlike the case for Charon, escape velocities on Pluto's smaller satellites are negligible compared with thermal speeds. They intercept Pluto's escaping atmosphere only in proportion to their geometric cross sections, with no enhancement effect from $CH_4$ molecules ballistically hopping around their surfaces. We consider Nix, since it exhibits a prominent red spot[29], though similar calculations could be done for any of the small satellites. Nix's effective radius is 20 km and its orbital distance is 48,760 km, so it intercepts $1.7 \times 10^{-7}$ of Pluto's escaping $CH_4$, or $8 \times 10^{18}$ molecules $s^{-1}$. Nix's escape velocity is roughly 10 to 20 m $s^{-1}$, so only $CH_4$ molecules that happen to hit an extremely cold region on their first impact will stick. Assuming Nix has a long-duration cold pole like Charon does (it might not, if its pole precesses rapidly), polar orientation with respect to radially outflowing gas from Pluto would diminish the accumulation rate by another factor of 4, leading to an accumulation rate of $\sim 1 \times 10^{-7}$ µm per year, about 20,000× slower than the accumulation rate at Charon's cold pole. Such slow accumulation might not be competitive with impact erosion of Nix's surface.

## Additional references

30. Hapke, B. *Theory of Reflectance and Emittance Spectroscopy*, Cambridge University Press, New York (1993).

31. Buie, M. W., Grundy, W. M., Young, E. F., Young, L. A., Stern S. A. Pluto and Charon with the Hubble Space Telescope I. Monitoring global change and improved surface properties from light curves. *Astron. J.* **139,** 1117-1127 (2010).

32. Buratti, B. J., et al. Global albedos of Pluto and Charon from LORRI New Horizons observations. *Icarus* (submitted; available at http://arxiv.org/abs/1604.06129).

33. Ross, R. G., Kargel, J. S. Thermal conductivity of solar system ices, with special reference to Martian polar caps. In: B. Schmitt, C. de Bergh, M. Festou (Eds.), *Solar System Ices*, Kluwer Academic Publishers, Boston, 33-62 (1998).

34. McKinnon, W. B., et al. The Pluto-Charon system revealed: Geophysics, activity, and origins. 47[th] Lunar and Planetary Science Conference, March 21-25, contribution #1995

   

## Code availability

   New Horizons MVIC and LORRI images were processed using the United States Geological Survey's Integrated Software for Images and Spectrometers (ISIS), available at https://isis.astrogeology.usgs.gov.  Thermal models were based on J.R. Spencer's thermprojrs.pro, available at https://www.boulder.swri.edu/~spencer/thermprojrs.  Hapke reflectance models were based on M.W. Buie's bidr2.pro, available at http://www.boulder.swri.edu/~buie/idl/pro/bidr2.html.

## New Horizons Science Team

S. A. Stern[1], F. Bagenal[2], K. Ennico[3], G. R. Gladstone[4], W. M. Grundy[5], W. B. McKinnon[6], J. M. Moore[3], C. B. Olkin[1], J. R. Spencer[1], H. A. Weaver[7], L. A. Young[1], T. Andert[8], O. Barnouin[7], R. A. Beyer[3], R. P. Binzel[9], M. Bird[10], V. J. Bray[11], M. Brozović[12], M. W. Buie[1], B. J. Buratti[12], A. F. Cheng[7], J. C. Cook[1], D. P. Cruikshank[3], C. M. Dalle Ore[13,3], A. M. Earle[9], H. A. Elliott[4], T. K. Greathouse[4], M. Hahn[14], D. P. Hamilton[15], M. E. Hill[7], D. P. Hinson[13], J. Hofgartner[12], M. Horányi[2], A. D. Howard[16], C. J. A. Howett[1], D. E. Jennings[17], J. A. Kammer[1], P. Kollmann[7], T. R. Lauer[18], P. Lavvas[19], I. R. Linscott[20], C. M. Lisse[7], A. W. Lunsford[17], D. J. McComas[4], R. L. McNutt Jr.[7], M. Mutchler[21], F. Nimmo[22], J. I. Nunez[7], M. Paetzold[14], A. H. Parker[1], J. Wm. Parker[1], S. Philippe[23], M. Piquette[2], S. B. Porter[1], S. Protopapa[15], E. Quirico[23],

H. J. Reitsema[1], D. C. Reuter[17], S. J. Robbins[1], J. H. Roberts[7], K. Runyon[7], P. M. Schenk[24], E. Schindhelm[1], B. Schmitt[23], M. R. Showalter[13], K. N. Singer[1], J. A. Stansberry[21], A. J. Steffl[1], D. F. Strobel[25], T. Stryk[26], M. E. Summers[27], J. R. Szalay[2], H. B. Throop[28], C. C. C. Tsang[1], G. L. Tyler[20], O. M. Umurhan[3], A. J. Verbiscer[16], M. H. Versteeg[4], G. E. Weigle II[4], O. L. White[3], W. W. Woods[20], E. F. Young[1], and A. M. Zangari[1].

[1]Southwest Research Institute, Boulder CO, [2]University of Colorado, Boulder CO, [3]NASA Ames Research Center, Moffett Field CA, [4]Southwest Research Institute, San Antonio TX, [5]Lowell Observatory, Flagstaff AZ, [6]Washington University in St. Louis, St. Louis MO, [7]Johns Hopkins University Applied Physics Laboratory, Columbia MD, [8]Universität der Bundeswehr München, Neubiberg 85577, Germany, [9]Massachussetts Institute of Technology, Cambridge MA, [10]University of Bonn, Bonn D-53113, Germany, [11]University of Arizona Lunar and Planetary Laboratory, Tucson AZ, [12]NASA Jet Propulsion Laboratory, La Cañada Flintridge CA, [13]Carl Sagan Center at the SETI Institute, Mountain View CA, [14]Rheinisches Institut für Umweltforschung an der Universität zu Köln, Cologne 50931, Germany, [15]University of Maryland, College Park MD, [16]University of Virginia, Charlottesville VA, [17]NASA Goddard Space Flight Center, Greenbelt MD, [18]National Optical Astronomy Observatory, Tucson AZ, [19]Université de Reims Champagne-Ardenne, 51687 Reims, France, [20]Stanford University, Stanford CA, [21]Space Telescope Science Institute, Baltimore MD, [22]University of California, Santa Cruz CA, [23]Université Grenoble Alpes, CNRS, IPAG, Grenoble France, [24]Lunar and Planetary Institute, Houston TX, [25]Johns Hopkins University, Baltimore MD, [26]Roane State Community College, Jamestown TN, [27]George Mason University, Fairfax VA, [28]Planetary Science Institute, Tucson AZ.

# Extended Data

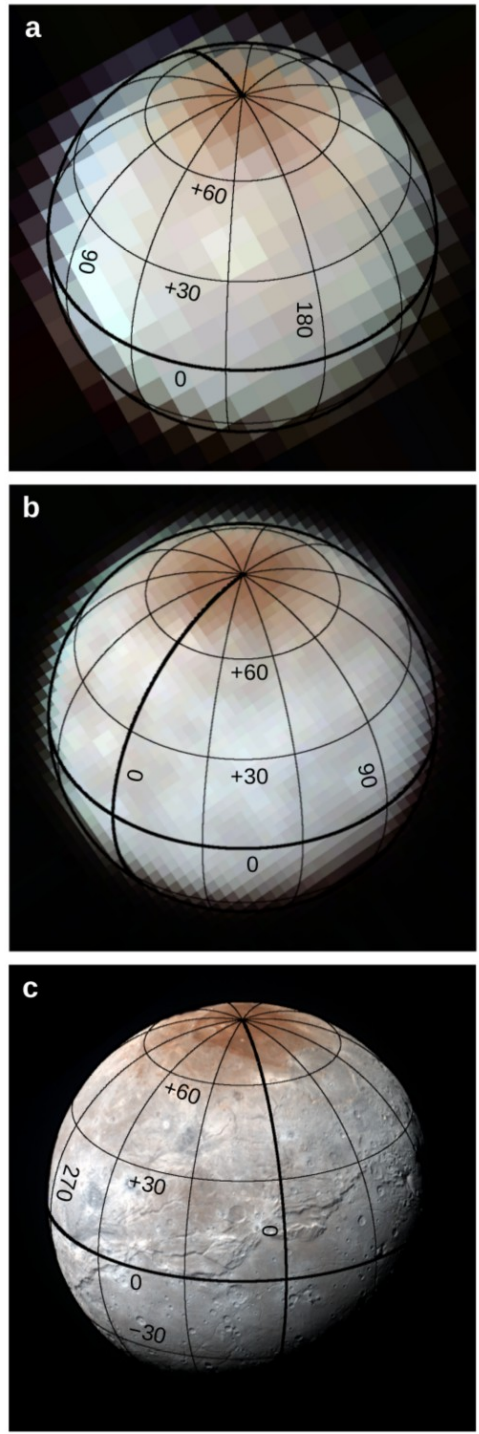

**Figure E1. MVIC colour images showing Charon's northern pole at three different sub-spacecraft central longitudes. a.** Observation obtained 2015 July 11 3:35 UT, with MET label 0298891582. **b.** Observation obtained 2015 July 13 3:38 UT, with MET label 0299064592. **c.** The same observation as Fig. 1a of the main text, obtained 2015 July 14 at 10:42 UT, with MET label 0299176432. Unlike in Fig. 1a, these images are not re-projected and not divided by photometric models. North, as defined by the angular momentum vector, is up.

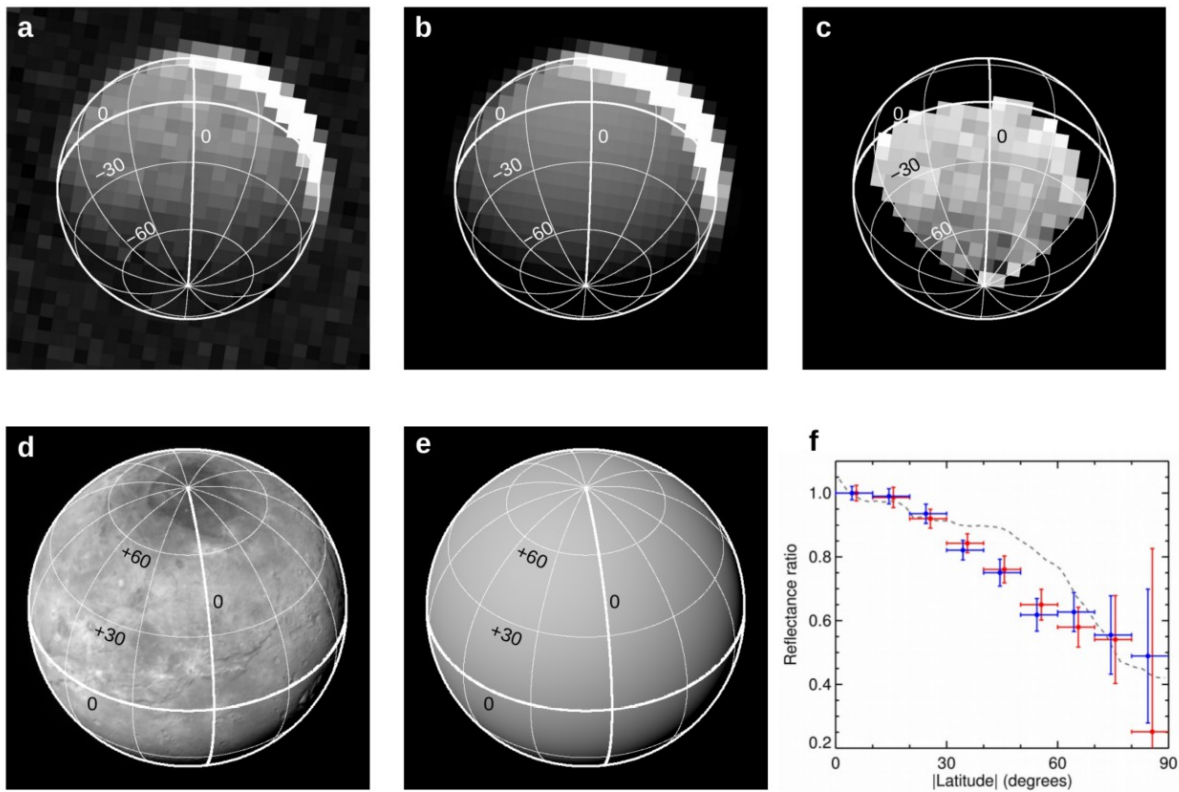

**Figure E2. Additional panchromatic observations of Charon's poles. a.** Second stack of 120 images of Charon's southern hemisphere illuminated by Pluto-shine obtained ~2 hours after the stack in Fig. 2a. **b.** Corresponding photometric model. **c.** Observation/model ratio. **d.** Sunlit northern hemisphere. **e.** Corresponding photometric model. **f.** Similar to Fig. 2d, with the first Pluto-shine stack indicated by blue points and the second stack indicated by red points (offset left and right for clarity). Horizontal bars indicate latitude bin width and vertical bars indicate the standard deviation of the mean within each latitude bin.

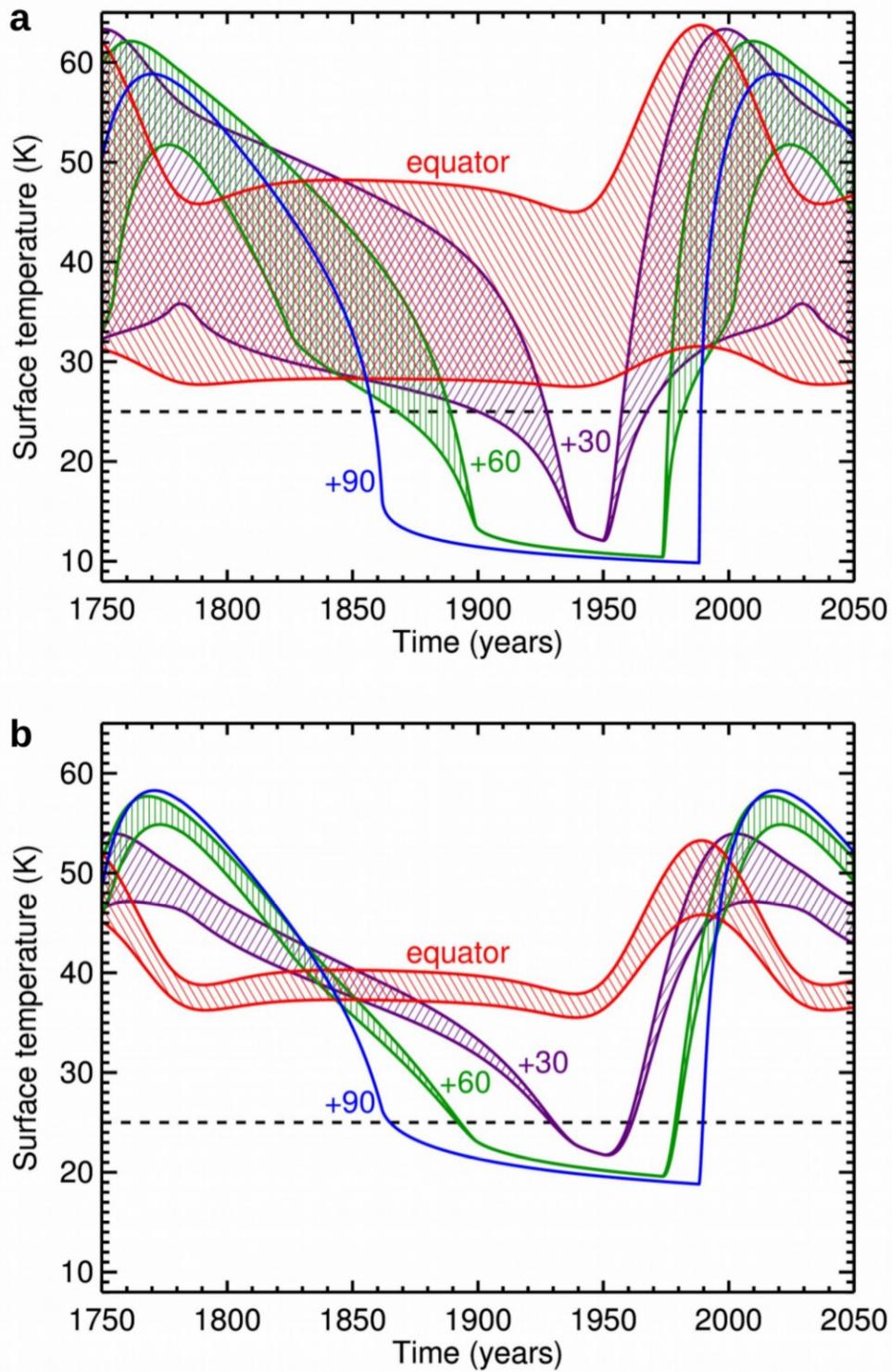

**Figure E3. Thermal models for cases of low and high thermal inertia. a.** Thermal history for lower limit thermal inertia $\Gamma = 2.5$ J m$^{-2}$ K$^{-1}$ s$^{-½}$. **b.** Thermal history for upper limit thermal inertia $\Gamma = 40$ J m$^{-2}$ K$^{-1}$ s$^{-½}$.